# Acceptance rates in *Physical Review Letters*: No seasonal bias

Are editorial decisions biased? A recent discussion in *Learned Publishing* [1-4] has focused on one aspect of potential bias in editorial decisions, namely seasonal (e.g., monthly) variations in acceptance rates of research journals. In four papers, data from six journals have been presented: *Psychological Science* (PS) and *Personality and Social Psychology Bulletin* (PSPB) by Shalvi et al. [1]; two anonymized journals by Hartley [2]; *Angewandte Chemie International Edition* (AC-IE) by Bornmann and Daniel [3]; and EPL (*Europhysics Letters*) by Schreiber [4]. In four out of these six journals, (PSPB, Hartley's two anonymized journals, and AC-IE), no statistically significant seasonal variation in acceptance rates was reported by the authors. For the journal PS, Shalvi et al. reported a discrepancy between monthly submissions and acceptances, an effect that was later criticized by Hartley [2]. Finally, in the EPL study Michael Schreiber expressed concern about comparatively high acceptance rates for July-submitted papers.

In this letter, we contribute to the discussion by analyzing data from *Physical Review Letters* (PRL), a journal published by the American Physical Society.

We looked at the 190,106 papers submitted to PRL in the 273-month period from January 1990 through September 2012. Only research articles are considered; Errata, Comments, Replies, Publishers' Notes, and Editorials, are excluded from the study. We sort papers by their submission date. Thus, the acceptance rate for, e.g., January 1990, is the fraction of papers submitted in that month that were ultimately published. Likewise, the number of published papers for the year 2000 refers to papers submitted in 2000 that were ultimately published, irrespective of when.

As we see from Figure 1, there is considerable seasonal variation in submissions within each year, with July being typically the month of most submissions. From 1990 up until 2008, the yearly trend of submissions is upward. In 2009, the *Physical Review Letters* editors decided to reinvigorate the journal's quality standards [5]. The plan was to 'raise the bar' by becoming more selective yet without drastically changing the scope of the journal [6]. The reinvigoration of the journal's standards appears to have had a direct effect on submissions that fell somewhat during the next two years, 2009 and 2010. By 2011 the upward trend in submissions had returned and it continued into 2012.

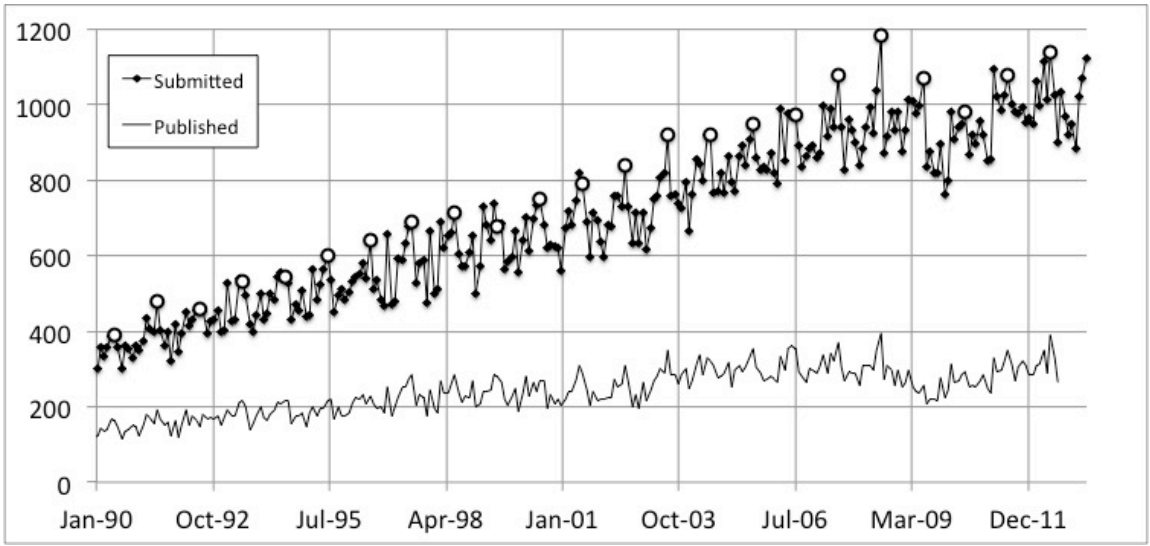

Fig. 1. Number of monthly submissions (upper curve) and monthly publications (lower curve), sorted by submission date, from January 1990 to May 2013. July submissions are shown in circles. No publication data are shown after September 2012 to avoid confusion with transient effects (papers in the pipeline whose fate had not been decided while this manuscript was in preparation).

In Figure 2 we plot the monthly acceptance rate $r_m$ versus the month $m$. We observe a small-scale, month-by-month, pattern, whereby $r_m$ appears to vary randomly within each year; and a large-scale, year-by-year, pattern, whereby the acceptance rate gradually decreases over time, accentuated by periods of increased overall rates as in 2003-2004 or 2010-2012.

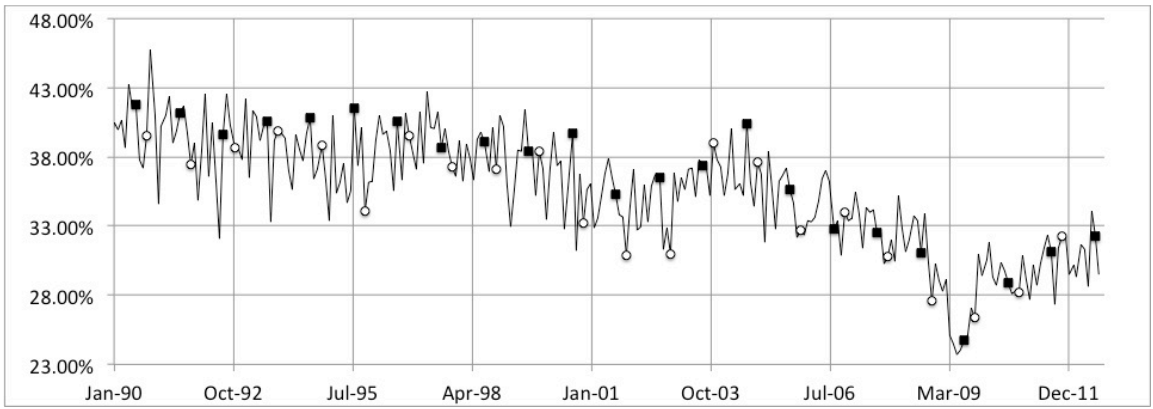

Fig. 2. Actual monthly acceptance rate $r_m$ for PRL, January 1990 – September 2012. Data for August are shown in squares, for November in circles.

In Table 1 we compute the *average* monthly acceptance rates $\langle r \rangle_m$ incorporating data from January 1990 to June 2012. (The average acceptance rate for, say, January is $\langle r \rangle_{Jan} = \sum_{Jan90}^{Jan12} r_m /23$.) To a good approximation, monthly acceptance rates are normally distributed. The standard deviation is $\sigma = 0.60\%$, and the standard deviation on the mean is $\sigma_m = 0.17\%$. In Figure 3 we plot the data from Table 1; the size of the error bar on each data point is $\pm\sigma$ on either direction. The mean (overall acceptance rate for all months and years, equal to 35.53%) is shown by the horizontal line.

| Month | Jan | Feb | Mar | Apr | May | Jun | Jul | Aug | Sep | Oct | Nov | Dec | Mean |
|---|---|---|---|---|---|---|---|---|---|---|---|---|---|
| average acceptance rate | 34.92 | 35.63 | 35.47 | 35.72 | 35.98 | 35.63 | 36.09 | 36.54 | 34.58 | 35.08 | 34.74 | 36.01 | 35.53 |
| Distance from the mean (in σ) | -1.02 | 0.17 | -0.10 | 0.32 | 0.75 | 0.17 | 0.94 | 1.69 | -1.59 | -0.75 | -1.32 | 0.80 | |

Table 1. Averaged monthly acceptance rates $\langle r \rangle_m$, for the period 1990-2012 (first 9 months of 2012 only).

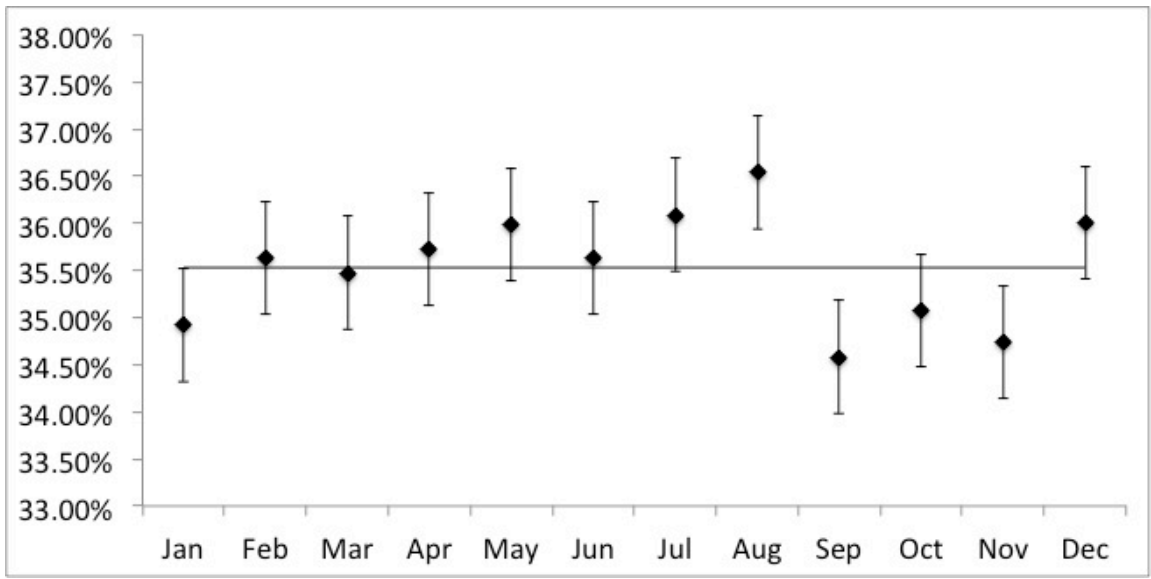

Fig. 3. Graphical representation of the data of Table 1. Averaged monthly acceptance rates $\langle r \rangle_m$, for the period 1990-2012 (first 9 months of 2012 only). The horizontal line denotes the mean acceptance rate for all months. For 12 points as in the plot, we expect from statistics four error bars not to touch the horizontal line, and all points to lie within $3\sigma$ from the line. Here, all points lie actually within $2\sigma$ from the line. So there are no deviations from the mean other than statistical fluctuations.

It is evident from Table 1 and Figure 3 that all twelve data-points corresponding to the monthly average acceptance rates lie within $2\sigma$ of the mean, while eight points lie within $\sigma$ of the mean. We observe no deviations from the behavior that would be expected due to random errors (see caption in Fig. 3).

To sum up, we have studied the 190,106 papers submitted to PRL from January 1990 until September 2012. No statistically significant variations were found in the monthly acceptance rates. We conclude that the time of year that the authors of a paper submit their work to PRL has no effect on the fate of the paper through the review process.

Are these results surprising? We think not. Since a typical PRL paper takes a few months to be published, the review process samples a several-month period thereby averaging out any possible differences in monthly acceptance rates. This smoothening is further enhanced by the sheer size of the journal (more than 1000 monthly submissions at present) and the fact that PRL largely operates with full-time in-house editors whose operation is less susceptible to the ebb and flow of academic schedules.

What about yearly acceptance rates? These may indeed vary with time in non-statistical (i.e., non-random) ways for a number of reasons, reflecting changes in a journal's editorial policies, the perception of a journal's prestige or popularity in the community, etc. Just for completeness, we show in Figure 4 the yearly-averaged acceptance rates, $\langle R \rangle_y$. We do observe an overall downward trend. The discussion of this trend and its causes goes beyond the scope of this letter.

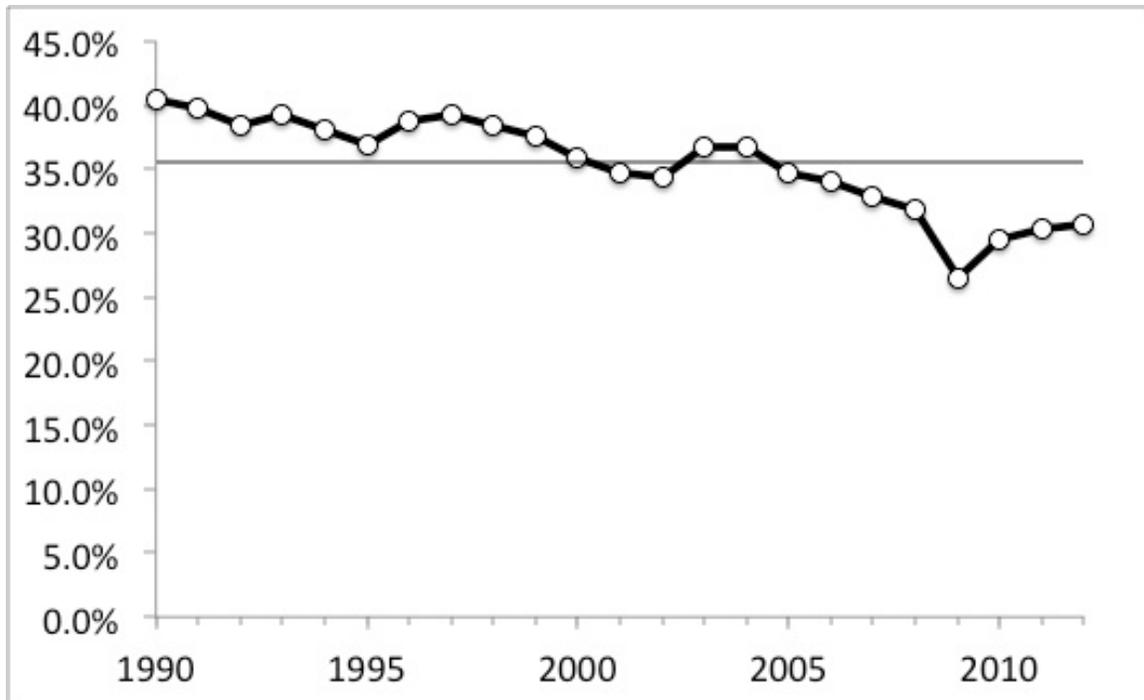

Fig. 4. Averaged yearly acceptance rate $\langle R \rangle_y$, for the period 1990-2012 (first 9 months of 2012 only).

Acknowledgments: I am grateful to my colleague Alex Klironomos, Senior Assistant Editor in *Physical Review B*, for stimulating discussions; and to Gene Sprouse, Editor-in-Chief of the American Physical Society, for encouragement.

Manolis Antonoyiannakis

Phone: +1-631-591-4076

[manolis@aps.org](manolis@aps.org)

Senior Assistant Editor, *Physical Review Letters*, The American Physical Society, 1 Research Road, Ridge, NY 11961

&

Adjunct Associate Research Scientist, *Department of Applied Physics and Applied Mathematics*, Columbia University, 500 West 120th St., New York, NY 10027